\pdfoutput=1

\documentclass[journal]{IEEEtran}
\usepackage{cite}
\usepackage[numbers,sort&compress]{natbib}

\ifCLASSINFOpdf
   \usepackage[pdftex]{graphicx}
\else
   \usepackage[dvips]{graphicx}
\fi

\usepackage{graphicx}
\usepackage{amsmath}
\usepackage{amssymb}   
\usepackage{array}
\newcolumntype{C}[1]{>{\centering\arraybackslash}m{#1}}
\usepackage{url}

\begin{document}
%
\title{Learning Analytics from Spoken Discussion Dialogs in Flipped Classroom}
%
%
%

\author{Hang~Su,
        Borislav~Dzodzo,
        Changlun~Li,
        Danyang~Zhao,
        Hao~Geng,
        Yunxiang~Li,
        Sidharth~Jaggi,
        and~Helen~Meng
        \thanks{Authors are with The Chinese University of Hong Kong, Shatin, N.T., Hong Kong }
}

\maketitle

\begin{abstract}
The flipped classroom is a new pedagogical strategy that has been gaining increasing importance recently. Spoken discussion dialog commonly occurs in flipped classroom, which embeds rich information indicating processes and progression of students' learning. This study focuses on learning analytics from spoken discussion dialog in the flipped classroom, which aims to collect and analyze the discussion dialogs in flipped classroom in order to get to know group learning processes and outcomes. We have recently transformed a course using the flipped classroom strategy, where students watched video-recorded lectures at home prior to group-based problem-solving discussions in class. The in-class group discussions were recorded throughout the semester and then transcribed manually. After features are extracted from the dialogs by multiple tools and customized processing techniques, we performed statistical analyses to explore the indicators that are related to the group learning outcomes from face-to-face discussion dialogs in the flipped classroom. Then, machine learning algorithms are applied to the indicators in order to predict the group learning outcome as High, Mid or Low. The best prediction accuracy reaches 78.9\%, which demonstrates the feasibility of achieving automatic learning outcome prediction from group discussion dialog in flipped classroom. 

\end{abstract}

\begin{IEEEkeywords}
Automatic Learning Outcome Prediction, Flipped Classroom, Learning Analytics, Spoken Discussion Dialog 
\end{IEEEkeywords}

%
\IEEEpeerreviewmaketitle

\vspace{-0.3cm}

\section{Introduction}
\label{intro}
\IEEEPARstart{L}{earning} analytics is concerned with collection and analyses of data related to learning in order to inform and improve the learning process or their outcomes \cite{siemens2011open}. Applying properly learning analytics can not only track student progress but also improve student performance \cite{olmos2012learning}.  Recent advancements in the development of data science and machine learning techniques has led to a rise in popularity of learning analytics within the educational research field. 

The flipped classroom is a new pedagogical method, which utilizes asynchronous video lectures and basic practice as homework, and conducts group-based problem solving discussions or activities in the classroom \cite{bishop2013flipped}. Since flipped classroom promotes cooperative learning \cite{de2017mathematics,wang2018investigation} and increases student engagement and motivation \cite{elmaadaway2018effects, long2017use}, it is gaining increasing importance for teaching and learning in recent years. A common in-class activity for the flipped classroom is student group discussions, where participants are involved in solving problems together. Such discussion dialogs embed rich information that cannot be captured objectively by conventional data, such as students' in-class sentiments, degree of concentration, amount of information exchange... etc. The information from in-class discussion dialogs may reflect the processes and progression of learning. Therefore, spoken discussion dialogs in flipped classroom deserve greater attention for learning analytics, which aims to collect and analyze the discussion dialogs in flipped classroom in order to explore indicators that reflect group learning outcomes. However, current studies in flipped classroom research field have not paid sufficient attention to this research problem, not only because of the difficulty in collecting such data in formal flipped classroom, but also because of the technological difficulties in analyzing such data. For one, speech recordings of group discussions present a source separation problem from multiple speakers, tackling overlapping speech, ambient noise and many other challenges. One may consider temporarily circumventing this problem of automatically transcribing the speech recordings by using manual transcripts, but this may be a laborious process. Moreover, many fields of technology including speech signal processing, natural language processing, data science and machine learning are required to be integrated together to conduct a holistic research, which is also a technical challenge. This work attempts to apply current technologies to analyze student group discussion dialogs in order to extract indicators of group learning outcomes. This may pave the way for deeper analysis into flipped classroom activities and their pedagogical values, and perhaps inform possible directions in developing future intelligent classroom.

\begin{figure}[!t]
\centering
\includegraphics[width=3.45in]{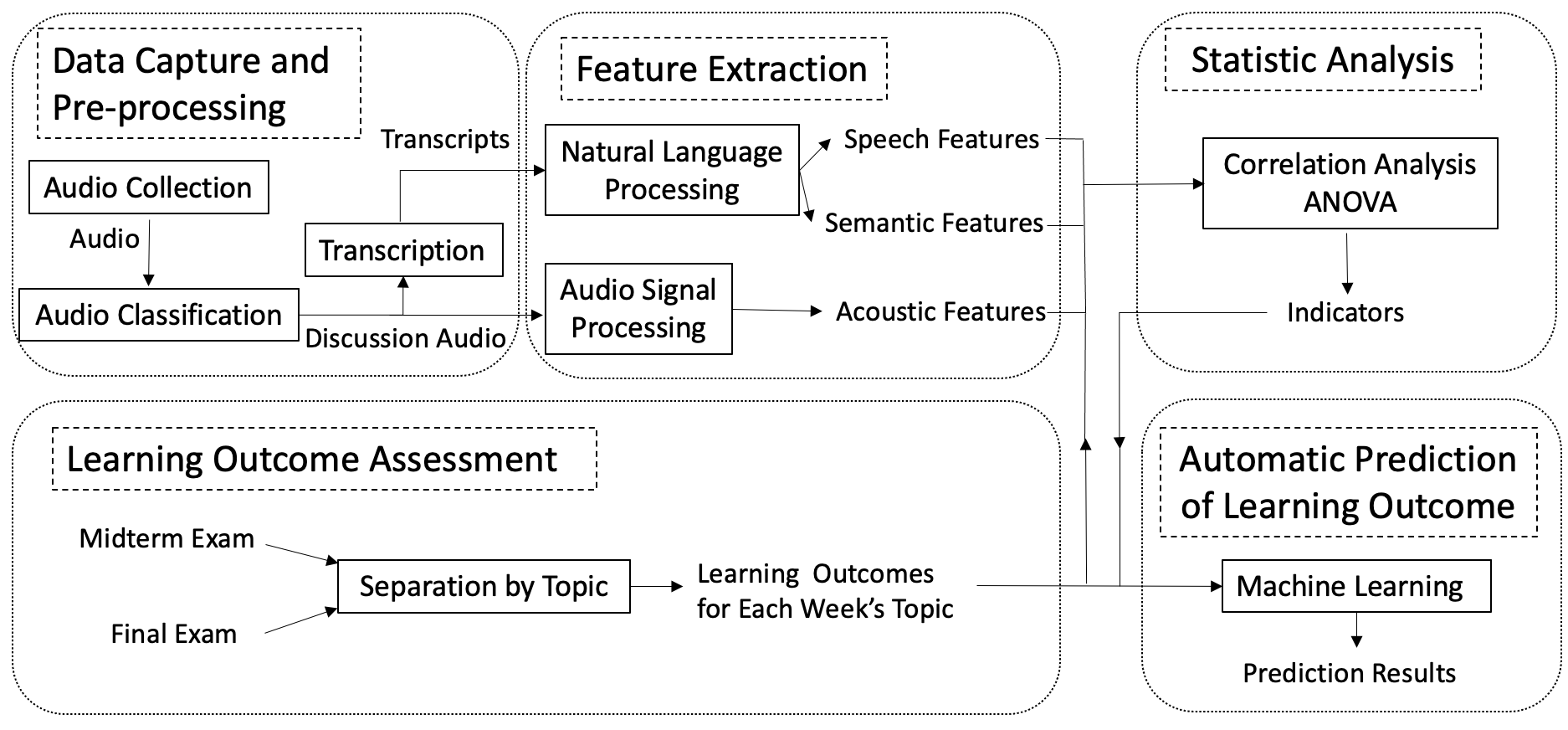}
\vspace{-0.5cm}
\caption{Framework of investigation}
\label{overview}
\vspace{-0.3cm}
\end{figure}

Fig. \ref{overview} shows the framework of this study. We have recently transformed a freshman engineering mathematics course from the conventional instructional strategy to the flipped classroom strategy. Students watched video-recorded lectures at home prior to group-based problem-solving discussion in class. Multiple audio data streams from multiple groups are collected non-intrusively throughout the semester, and a customized speech classification technology is applied to obtain student group discussion dialog audios. The dialog audios are then manually transcribed. Then, spoken dialog features are extracted by using some proper speech and language processing tools and techniques from bilingual transcription text data and audio data. Learning outcome is measured in terms of examinations. Several essential indicators from discussion dialogs that reflect the group learning outcome are found by statistical analysis. Then, indicators obtained from statistical analysis are used as input to a variety of machine learning algorithms in order to predict the group learning outcome as High, Mid or Low. Results indicate that it is feasible to use the indicators we found to automatically predict group learning outcome from face-to-face discussion dialog in flipped classroom.

To the best of our knowledge, this is a novel work in investigating learning analytics from spoken discussion dialogs in the flipped classroom. Contributions of the study include:
\begin{itemize}
    \item A new Chinese and English code-switched flipped classroom discussion audio corpus were non-intrusively recorded and transcribed.
    
    \item  Proper tools and customized techniques were introduced to obtain the spoken dialog features from multi-modal and code-switch discussion data, which not only could be generalized to other discussion dialogs, but also demonstrated the feasibility of dealing such data for learning analytics.


    \item Indicators of group learning outcome were found out from flipped classroom in-class discussion dialog.

    \item  The feasibility of predicting group learning outcome from discussion dialog in flipped classroom was demonstrated.

\end{itemize}  

The remaining sections of the paper are organized as follows:
The related work of this research is introduced in Section 2. The capture and pre-processing of flipped classroom discussion dialog corpus is illustrated in Section 3. The spoken dialog features extraction from the audio and text data is described in Section 4. The group learning outcome assessment, as well as exploration of indicators that reflect the group learning outcome from in-class spoken dialogs of flipped classes are described in Section 5. The automatic prediction of group learning outcome based on flipped classroom in-class discussion dialog features is described in Section 6. The conclusion is stated in Section 7.

\vspace{-0.1cm}

\section{Related Work}
\label{relate}
\subsection{Quantitative Research of Flipped Classroom}
With the popularity of flipped classrooms in practical teaching and learning, much research in the field of education has been devoted to this topic in recent years.

Much previous work aimed to investigate whether the flipped classroom enhances student learning and motivation, often through comparisons with traditional teaching approaches. [Hew and Lo 2018] claimed that the flipped classroom approach in health professions education yields a significant improvement in student learning compared with traditional teaching methods \cite{hew2018flipped}.  By measuring students' performance on the final examination and students' self-reported satisfaction, [Joseph et al. 2021] found that flipped classroom improves nursing students' performance and satisfaction in anatomy and physiology compared to a traditional classroom \cite{joseph2021flipped}. [Zhao et al. 2021] used questionnaires to demonstrate that students in the flipped class have a better learning performance than students in a traditional classroom \cite{zhao2021innovative}. By conducting a self-efficacy survey, [Namaziandost and Çakmak 2020] found that students in a flipped classroom have higher self-efficacy score compared to traditional classroom \cite{namaziandost2020account}.

The strategy and design of the flipped classroom approach are important for its effectiveness \cite{milman2012flipped, awidi2019impact}. [Chen and Yen 2021] investigated how to make a better pre-class instruction when using animated demonstrations, in order to enhance student engagement and understanding in the flipped classroom \cite{chen2021learner}. By using self-efficacy questionnaire and interview questions, [Hsia and Sung 2020] found that the flipped classroom with peer review could significantly enhance students' intrinsic motivation and strengthen their focus and reflection during activities \cite{hsia2020effects}. [Song and Kapur 2017] compared the “productive failure-based flipped classroom”(PFFC) pedagogical design with traditional flipped classroom pedagogical design and they found that students in PFFC achieved higher scores in solving conceptual questions \cite{song2017flip}. [Lin et al. 2021] proposed the “Scaffolding, Questioning, Interflow, Reflection and Comparison” mobile flipped learning approach (SQIRC-based mobile flipped learning approach) and found that the proposed approach improved the students' learning performance, self-efficacy and learning motivation compared to traditional flipped classroom \cite{lin2021promoting}.


Previous work have also studied the impact of different indicators on learning performance in flipped classrooms. The indicators of students' learning performance are investigated based on students' learning strategies measured from students' behaviors in pre-class activities such as online course and online assessment \cite{jovanovic2017learning, jovanovic2019predictive}. [Wang 2021] explored indicators of learning outcome in the flipped classroom by analyzing learning management systems log data and questionnaires \cite{wang2021interpreting}. [Murillo-Zamorano et al. 2019] used questionnaires to investigate the causal relationships of knowledge, skills, and engagement with students' satisfaction \cite{murillo2019flipped}. [Zheng and Zhang 2020] did analyses on students' response of survey and found that students' behaviors such as peer learning and help seeking were positively associated with students' learning outcomes \cite{zheng2020self}. [Lin and Hwang 2018] studied how the students' online feedback on video learning materials and video recordings of classmates' work was relevant to their performance \cite{lin2018learning}. Quantitative measures such as learning hours and attendance have also been studied to identify their relevance to learning performance \cite{boeve2017implementing}.


Although many quantitative researches on flipped classrooms have been conducted, few research paid attention to the in-class discussion dialogs. Spoken discussion dialogs commonly occur in flipped classrooms, which embed rich information related to learning that cannot be provided objectively by conventional data. Therefore, the spoken discussion dialogs in flipped classrooms should not be ignored in learning analytics.

\vspace{-0.1cm}

\subsection{Research on Student's Discussion Analyses}
Due to increasing availability of sensors as well as computational resources and algorithms, some studies on analysis of face-to-face discussion were performed. [Kubasova et al. 2019] used verbal and nonverbal features of group discussions for predicting group performance in a cooperative game \cite{kubasova2019analyzing}. [Avci and Oya 2016] used aural and visual cues with their novel classifier to predict the group performance in decision-making tasks \cite{avci2016predicting}. [Murray and Oertel 2018] predicted group performance in task-based interactions by utilizing speech and linguistic features \cite{murray2018predicting}. However, these studies were conducted in the context of gaming or decision-making rather than in an educational context. [Ochoa et al. 2013] used a data set of video, audio and pen strokes to extract features that can discriminate between experts and non-experts in math problem discussions \cite{ochoa2013expertise}. [Scherer et al. 2012] investigated predictors from audio and writing modalities for the separation and identification of socially dominant leaders and experts within a study group \cite{scherer2012multimodal}. [Martinez-Maldonado et al. 2013] determined the level of cooperation at an enriched interactive tabletop from verbal and physical information \cite{martinez2013capturing}. [Reilly and Schneider 2019] conducted an educational experiment which showed that Coh-metrix could extract learning gain indicators, and they also predicted the quality of collaboration in discussions \cite{reilly2019predicting}. [Spikol et al. 2017] estimated the success of small collaborative learning groups in an experimental setting that extracted features from vision, user generated content and learning objects \cite{spikol2017estimation}. To the best of our knowledge, there is no prior work that investigates how the learning dynamics captured from student group discussion dialogs correlates with the learning outcomes in the flipped classroom.

\vspace{-0.1cm}

\section{Discussion Dialog Data Capture and Pre-processing}
\label{Data}
\subsection{Flipping a Freshman Engineering Mathematics Course}
\label{course}
This research is conducted in an engineering mathematics flipped classroom \cite{jaggisystematic}, which has recently been transformed from the conventional instructional setup. In this course, students watch lecture videos before class and solve in-class exercise problems in groups, aided by the professor and teaching assistants during the class time. The quality of group discussions in such flipped classroom is highly related to student group grade since problem solving discussions take up most of the class time. The remaining, minor portion of class time is for announcements and sharing of good solutions to given problems. 

\vspace{-0.1cm}

\subsection{Audio Capture of In-Class Discussion Dialogs}
\label{audio-collection}
\begin{figure}[!t]
\centering
\vspace{-0.1cm}
\includegraphics[width=3.4in]{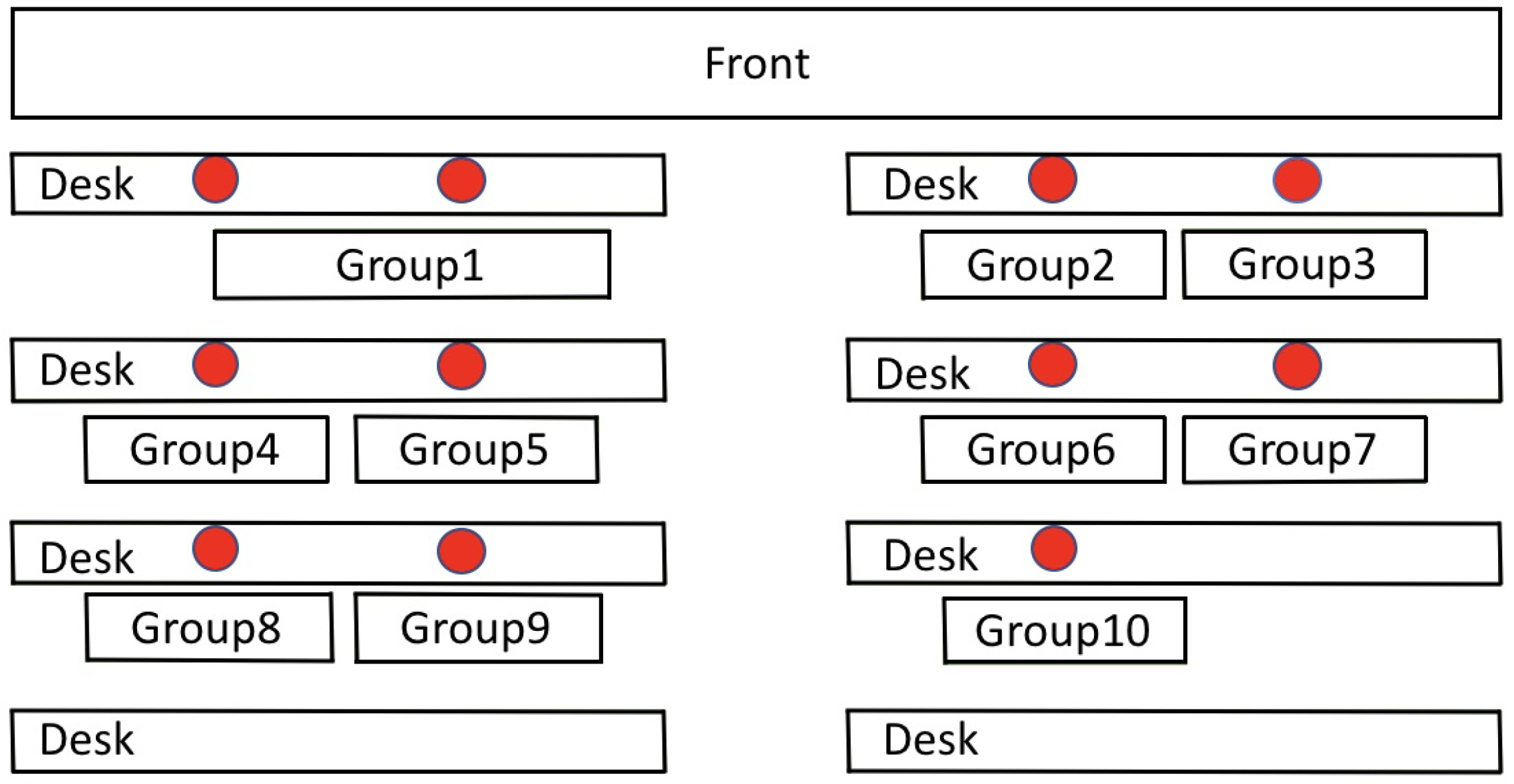}
\vspace{-0.2cm}
\caption{Recording environment}
\vspace{-0.3cm}
\label{environment}
\end{figure}

Students enrolled in the class form their own groups of 3-4 members for the semester.  Students in the same group sat together in the classroom, which was a laboratory with computers.  Fig. \ref{environment} shows the setup, where the red dots showed the placement of a speech recorder.  We had selected the TASCAM DR-05 recorder because of its relative small size, and hence they could be placed in a non-intrusive position for each student group.  Adjacent recorders were placed at least 1.5 meters apart and were positioned to point towards the members in the same student group for sound capture. The sampling rate was set at 44.1 kHZ.  We obtained consent from the students through a consent form in order to record their speech during in-class discussions.  We also offered the option that students were free to press the button to stop recording at any time during class, but no student did that throughout the whole semester. After the semester's add-drop deadline for courses, we recorded discussions from the ten groups in class until the end of the semester.

\begin{figure*}[!t]
\centering
\vspace{-0.2cm}
\includegraphics[width=6in]{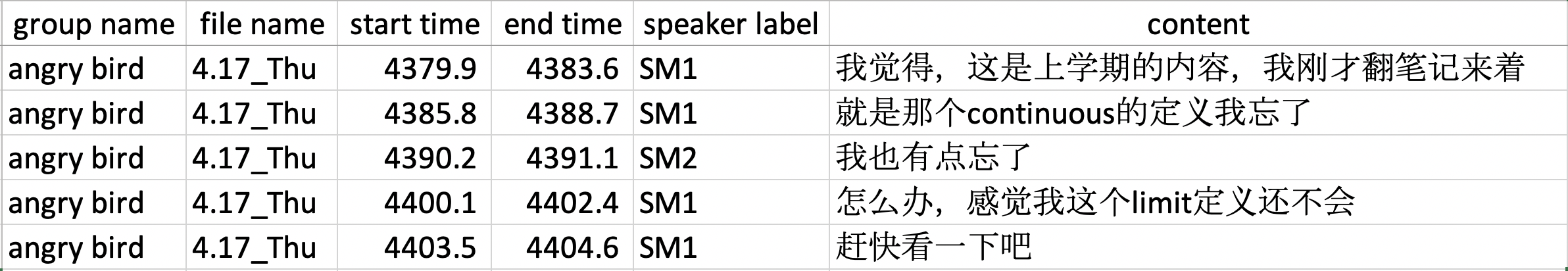}
\vspace{-0.2cm}
\caption{{Transcription example from group ``angry bird" in the class of 17 April. Speech time in second, speaker label, and speech content are recorded.}}
\vspace{-0.3cm}
\label{transcription example}    
\end{figure*}

\vspace{-0.1cm}

\subsection{Pre-processing of Audio Recordings}
\label{audio-content}
Each recording contains discussion speech as well as broadcasted information such as general class announcements and solutions to problems. Since we are more interested in the portions of the audio that only contain group discussions, an unsupervised classification of lecture and discussion is achieved by using a customized audio processing technique \cite{su2019unsupervised}. In designing the classification algorithm, we aim to fully leverage the simultaneous recordings from the devices placed around the classroom. Specifically, lectures have high similarity across various simultaneous recordings. Therefore, after recordings are coarsely synchronized to a common start time by using Fast Fourier Transform (FFT) convolution, we compute the normalized similarity between a given window and temporally proximate window segments in other recordings. Histogram plot automatically categorizes higher similarity windows as lecture and lower ones as discussion. With the proposed method, broadcasted information is eliminated in each of our audio, and only discussion speech is further analyzed in this work.

\vspace{-0.1cm}

\subsection{Discussion Dialog Transcription}
\label{Trans}
Presently, we do not have a sufficiently accurate, Chinese-English code-switched automatic speech recognition system that can transcribe the discussion audio recordings.  Instead, we use manual transcriptions, aiming for a diarization task of “who spoke what at when”. Fig. \ref{transcription example} shows an example of the transcription. Speech time information, speaker information and discussion content are recorded.

Discussion dialogs are segmented based on two criteria: 1) Speaker changes from one person to another. 2) The occurrence of a long pause of duration greater than a second. The start and end times of each segment are recorded. Transcribed speakers are classified as the professor, teaching assistants, in-group students and out-of-group students. The transcription focuses mainly on the speech of the in-group students as well as the professor, teaching assistants and out-of-group students who talk to in-group students. All other sounds are ignored. Each speaker is identified by a generic speaker label to protect privacy, such as Student Male 1(SM1), Student Female 1(SF1), Teaching assistant Male 1(TM1), Professor(P), Out-group Male 1(OM1), etc. Except for out-of-group speakers, the voice recordings of the in-group speakers are given to the transcriber so that they can label them with the most similar identifier. If a voice does not match any of the in-group speakers, it will be labeled as out-of-group.  Overlapping speech from multiple speakers may occur, and the transcribers will try their best to offer separate transcriptions for the two speakers.  Since this course is taught in English but students are mainly Chinese, the group discussions are often conducted in Chinese-English code-switch modes.  The transcriptions will strictly follow the languages spoken. All of the transcriptions are double checked to ensure the quality of the transcription. All transcribers signed a confidentiality agreement to protect the confidentiality of the recordings.

\vspace{-0.1cm}
\section{Dialog Features Extraction}
\label{Feat}
Multiple salient spoken dialog features should be extracted from the discussion dialog audio and text data for learning analytics. Speech features, semantic features and acoustic features are considered in this work. All of the features are initially extracted from each dialog segment, which is the basic unit of technical processing of dialog data. Then group-level features (e.g. sum/mean/variance of all segments) are created to represent the features of group discussion dialog (e.g. sum of math terms in discussion). Details of how to extract every feature from each dialog segment as well as the extraction of group-level features will be shown in this section.

\vspace{-0.18cm}
\subsection{Speech Features}
\label{speech}
In order to capture the activity and amount of information exchange in discussion, the following speech features are considered.

\begin{itemize}
    \item Speech Time (ST)
    
    The original audio data is segmented and annotated with start and end timestamps. The duration between the start time and end time is calculated as the speech time of that segment.

    \item Number of Words (NoW)
    
     The number of words for each segment is a measure that differs in English and Chinese. English is counted in word-unit and Chinese is counted in character-unit. The punctuation is ignored.

    \item Number of Turns (NoT)
    
    As mentioned in Sect.~\ref{Trans}, the discussion dialog are segmented if speaker changes or a long pause of duration greater than one second occurs. The number of dialog turns is counted as number of dialog segments in transcription.

    \item Speaking Rate Score (SRS)
  
     Speaking rate for one segment is initially calculated as the number of words divided by the speech time. Then each segment is rated as \textit{Slow}, \textit{Normal} or \textit{Fast}, which are subsequently scored as 0,1,2 respectively. The classification thresholds for \textit{Slow}, \textit{Normal} and \textit{Fast} are based on statistical findings of Yuan et al. \cite{yuan2006towards}, which claimed that the normal speed of short spoken English is 100-120 words per minute and the normal speed of short spoken Chinese is 225-255 characters per minute. Thus the thresholds are calculated as an equation (\ref{ThresSN})(\ref{ThresNF})
    \begin{equation}
    Thres_{S/N} = 100*Eng_{prop} + 225*Cn_{prop}
    \label{ThresSN}
    \end{equation}
    \begin{equation}
    Thres_{N/F} = 120*Eng_{prop} + 255*Cn_{prop}
    \label{ThresNF}
    \end{equation}
    where $Thres_{S/N}$ is the classification threshold between \textit{Slow} and \textit{Normal}, $Thres_{N/F}$ is the classification threshold between \textit{Normal} and \textit{Fast}, $Eng_{prop}$ and $Cn_{prop}$ are the proportion of English and Chinese used in one segment correspondingly.
    
\end{itemize}  

\vspace{-0.18cm}
\subsection{Semantic Features}
\label{semantic}
\subsubsection{Topic Relevance of Discussion}
\label{topic-sec}
In this work, some customized features are designed to measure the topic relevance of discussions by Natural Language Processing (NLP) techniques.

\begin{itemize}

    \item Math Terms (MT)
    
    Since discussions are conducted in an engineering mathematics flipped classroom, detecting the mathematical terms in students' discussion is conducive towards measuring topic relevance.  In this work, a customized bilingual math glossary is derived from engineering mathematics textbooks in both Chinese and English. A count of matching math terms in each segment is calculated based on the glossaries.
    
    \item Topic Relevance Score (TRS)
    
    The use of word matching may not sufficiently reflect the topic relevance of students' discussion, because there may be on-topic discussions that do not contain the exact terms in glossaries. Therefore, a scoring mechanism to measure the degree of topic relevance of each segment needs to be developed. We adopt word embeddings, a predominant word representation in NLP.  A word can be represented as a vector, and words with similar meaning are closer in the vector space. In this study, we design a framework to score the topic relevance of each segment based on the word embedding technique. The score is between zero and one, higher score indicates higher topic relevance.

\begin{figure*}[!t]
\centering
\vspace{-0.3cm}
\includegraphics[width=5in]{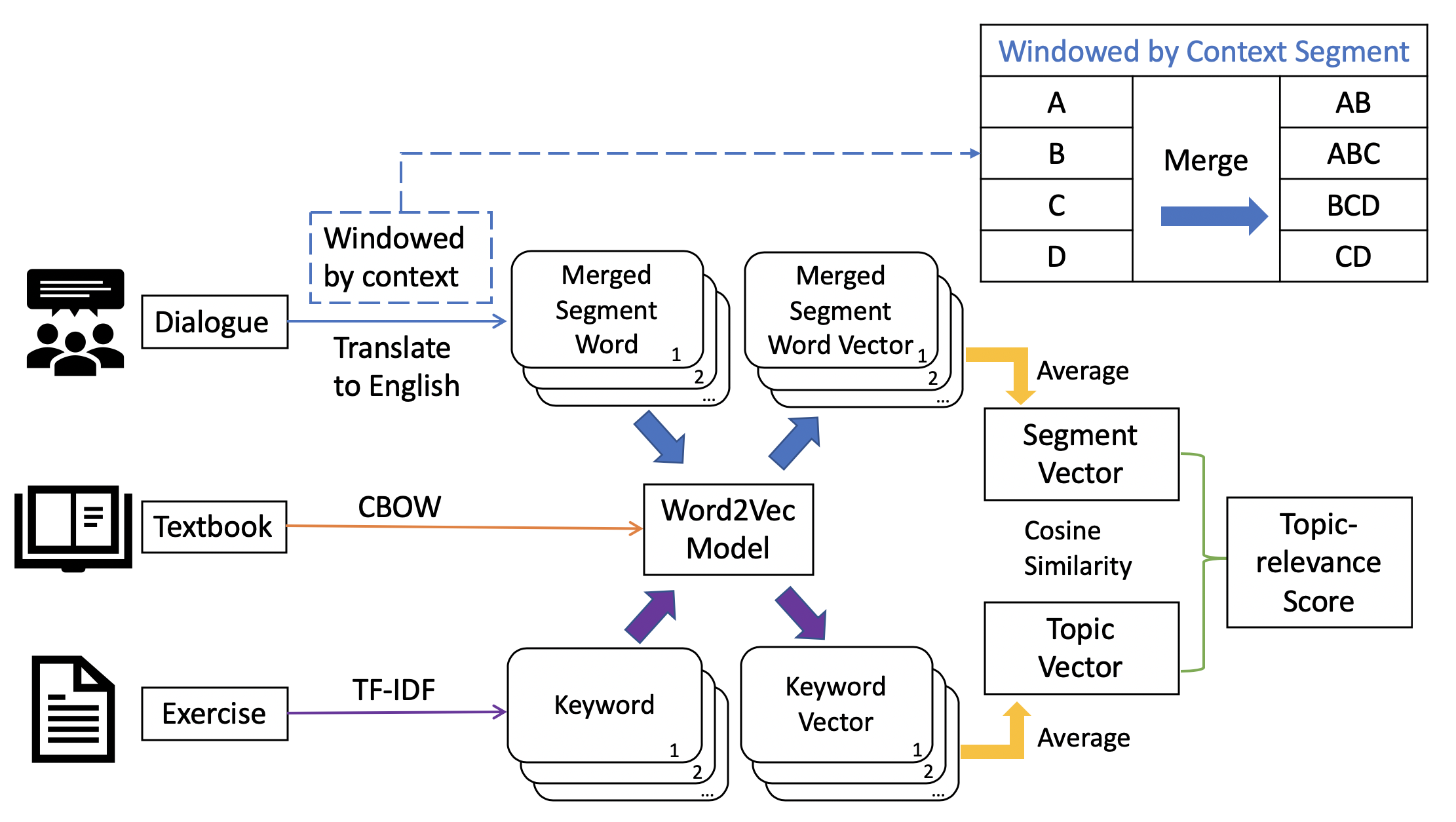} 
\vspace{-0.2cm}
\caption{{Framework of topic-relevance score measurement}}
\label{topic-relevance}    
\vspace{-0.3cm}
\end{figure*}

    Fig. \ref{topic-relevance} provides an illustration of the framework. To represent the words in the field of mathematics, we trained a word embedding model based on the textbook \cite{kreyszig11} of this course. We used the Continuous Bag Of Words (CBOW) model architecture for word-embedding training, which is implemented by the Gensim Word2Vec tool \cite{rehurek_lrec}. Since students mainly discuss the in-class exercises which are assigned with different topics every week, keywords from the in-class exercises can represent the class topic for each week. Therefore, keywords from in-class exercises of each week are extracted using the TF-IDF algorithm \cite{jones1972statistical}. Common NLP practice takes the average word embedding vectors to represent the general meaning of combined words \cite{kenter2016siamese,kenter2015short,zhao2015ecnu}.  Therefore, word vectors of keywords from each week's in-class exercises are averaged to get a \textit{topic vector} that represents the class topic of that week. In order to measure the topic relevance of each discussion segment, each segment should be represented by a vector which is the average word vector in that segment. Notably, instead of directly using each isolated segment, we consider the dialog context of neighbouring segments. The dialog context of a discussion segment further defines the meaning of that segment. Consider a simple scenario: student A said 'Yes it is' in a math-related discussion and student B said 'Yes it is' in a game-related discussion. Although the words are identical, the topic relevance scores should be different due to the different contexts. Therefore, a context window is used to pre-process the segments. As shown in Figure \ref{topic-relevance}, each segment is merged with one segment before and one segment after, so that the words in a merged segment can better represent the meaning of the original segment. The course textbook and exercises are written in English, while discussions are usually conducted in Chinese.  To standardize the use of language, segments are translated into English by the Google Translator API \cite{google-translation}. Then, the vectors of merged segment words are averaged to get the \textit{segment vector} which represents the meaning of that segment. Finally a cosine similarity between the \textit{topic vector} and \textit{segment vector} is calculated, and each segment finally has a topic relevance score.

\end{itemize}

\subsubsection{Context Cohesion of a Discussion}
\label{context}
Context cohesion is an important feature for discussion dialogs, which was also used in previous work on discussion analysis \cite{kubasova2019analyzing}. This study computes the cohesion score to reflect context cohesion:

\begin{itemize}
    \item Cohesion Score (CS)
    
     Cohesion score is calculated as a cosine similarity between current and subsequent segments in a discussion. The vector representation of each segment is the same as the vector representation used for topic relevance score.

\end{itemize}

\subsubsection{Social Aspects of the Dialog}
\label{social}
Social aspects of dialog, such as sentiment and cognitive processing, are also important features of discussion dialog. In this work, Linguistic Inquiry and Word Counting (LIWC) is applied to measure the social aspects of dialogs.

\begin{table}[!t]
\renewcommand{\arraystretch}{1.2}
\caption{LIWC categories and explanations}
\label{LIWC_table}
\centering
\begin{tabular}{p{1.1in}p{2.1in}}
\hline
Category & Explanation \\
\hline
Positive Emotion (PE) & An emotional reaction designed to express a positive affect \\
Negative Emotion (NE) & An unpleasant emotional reaction designed to express a negative affect \\
Anger & A strong feeling of annoyance, displeasure, or hostility \\
Anxiety & Feeling of worry, nervousness, or unease about something \\
Risk & The possibility that something unpleasant or unwelcoming will happen \\ 
Assent & The expression of approval or agreement \\
Negation & The contradiction or denial of something \\ 
Affect & Words depicting emotions  \\ 
Tentative (Tent) & Words that indicate intention to try something \\
Certainty (Cert)&  Having or showing complete conviction about something \\
Insight & An accurate and deep understanding of something  \\
Causation (Caus)& The relationship between cause and effect \\
Conjunction (Conj)&  A word used to connect clauses or sentences or to coordinate words in the same clause \\
Filler & A linguistic but meaningless unit that fills a particular slot in syntactic structure \\
Interrogative (Int)& Words used to purpose questions, e.g, who, what, where, how \\
Differentiation (Diff)& The action or process of distinguishing between two or more things \\
Comparison (Comp)& A consideration or estimate of the similarities or dissimilarities between two things \\ 
Quantity Unit (QU)& Measure word, usually occurs together with math-related words, more frequently used in Chinese \\  
Leisure & Use of free time for enjoyment \\
\hline
\end{tabular}
\vspace{-0.35cm}
\end{table}

\begin{itemize}
    \item LIWC Features (LIWC) 

    LIWC is a dictionary-based text analysis application, which provides an efficient and effective method for studying various emotional, cognitive, and structural components present in individuals' verbal and written speech samples \cite{huang2012development, pennebaker2015development}. The original LIWC provides hundreds of categories of vocabulary, but not all of the categories fit the current investigation. Therefore, we manually select the suitable categories in LIWC, which is shown in Table \ref{LIWC_table}. Positive Emotion (PE), Negative Emotion (NE), Anger, Anxiety, Risk, Assent, Negation and Affect are used to measure the sentiments of the dialog. Tentative (Tent), Certainty (Cert), Insight, Causation (Caus), Conjunction (Conj), Filler, Interrogative (Int), Differentiation (Diff) and Comparison (Comp) are used to indicate cognitive processing. Moreover, Quantity Unit (QU) and Leisure can be used as supplements to measure the degree of topic relevance of dialog. Furthermore, LIWC provides both Chinese and English dictionaries for analysis. In order to deal with the bilingual dataset, we process the English words and Chinese characters separately by using both English and Chinese version of LIWC dictionary. The number of matched words for each LIWC category is counted in each segment based on the bilingual LIWC dictionaries.

\end{itemize}

\vspace{-0.2cm}
\subsection{Acoustic Features}
\label{acoustic}
The audio signal also contains a variety of information related to the dialog, such as emotions during discussions. Therefore, we extract certain features directly from the audio data. All of the audio features are first calculated for every frame (0.01 second in duration). Then the mean, maximal and minimal values within a segment are calculated as feature values of each segment. Gender broadly affects the values of audio features, especially related to F0. Presently we are not investigating the gender of the speakers, hence, all the audio features are gender-normalized. The following features are considered in this work:

\begin{itemize}
    \item Fundamental Frequency (F0)
    
    The fundamental frequency (F0) of a speech signal refers to the approximate frequency of the (quasi-)periodic structure of voiced speech signals. F0 can be used to measure the level of arousal in discussions and it was used to disambiguate discussion roles between a leader and an expert in the work of Scherer et al \cite{scherer2012multimodal}. In this work, F0 detection is performed by COVAREP \cite{degottex2014covarep}.

    \item Energy
    
    Energy can be used to measure the level of arousal in discussions, and it was also used in the work of Scherer et al \cite{scherer2012multimodal}.
    Energy of each frame is measured by Equation (\ref{energy}),
    \begin{equation}
        Energy=\sum_{n}^{N} {x^{2}(n)} \label{energy} 
    \end{equation} 
    where $N$ is the total number of samples in one frame, $x$ represents the audio signal, $n$ is the current sample in audio signal.
    

    \item Formants
    
    In speech signal processing community, first formant (F1), second formant (F2) and third formant (F3) are the first, second and third acoustic harmonics of the fundamental frequency, respectively. Formants are important for emotion detection in Mandarin \cite{xu2014survey}, and they were also used in the work of Worsley and Blikstein \cite{worsley2010towards}. In our work, the value of F1, F2 and F3 are calculated based on the Praat software \cite{boersma2002praat}.

    \item Glottal Features
   
     Glottal features describe the movements of vocalizing muscles. Glottal features have been shown to have a relationship with emotions and demeanor \cite{tao2005features,wang2006,alku2002normalized,patel2011mapping}, and they have been used in the field of learning analytics \cite{scherer2012multimodal}. In this work, creak \cite{laver1980phonetic}, rd \cite{degottex2010phase, huber2012glottal}, Peak Slope (PS) \cite{kane2011identifying}, the difference between the first and second harmonics (H1-H2), Quasi Open Quotient (QOQ) \cite{hacki1989classification}, Maxima Dispersion Quotient (MDQ) \cite{kane2013wavelet} and Normalized Amplitude Quotient (NAQ) \cite{alku2002normalized} are extracted by using COVAREP \cite{degottex2014covarep}.


\end{itemize}

Audio feature abbreviations are defined and used as follows: \{Max, Min, Avg\}\_\{F0, Energy, NAQ, etc.\}. For instance, Max\_F0 represents that the feature value of one segment is obtained by taking the maximum value of F0 within that segment.

\vspace{-0.12cm}
\subsection{Derivation of Group-level Features}
\label{group}
The features mentioned in Sections~\ref{speech}, ~\ref{semantic} and ~\ref{acoustic} are created for each dialog segment. In order to derive some characteristics in group discussion dialog, all features need to be transformed to group-level. In this work, 7 types of group-level features in 2 aspects are considered.

The segments spoken by an in-group student can be aggregated together to be the feature of this student. Characteristics of a group discussion can be measured by the mean/variance of all in-group students' characteristics. Therefore, we first take the sum/mean of feature value across the segments for each in-group student, and then the mean/variance among all in-group students. For example, we first calculate the sum/mean number of math terms per segment for each in-group student, and then take the mean/variance across the students within the group. This provides us with 4 types of group-level features as shown below (take math term as example):

\begin{itemize}
   \item Mean of the total number of math terms spoken by each in-group student

   \item Variance in the total number of math terms spoken by each in-group student
   
   \item Mean of the average number of math terms spoken by each in-group student

   \item Variance in the average number of math terms spoken by each in-group student
    
\end{itemize}

In addition, we also look at the entire discussion dialog of a group for the topic of a given week, which contains dialog segments including those from the professor and teaching assistants to give guidance, as well as from students in a neighboring group who join in temporarily. We can compute the sum/mean/variance for features across all segments in the entire discussion. This provides us with 3 types of group-level features as presented below (take math term as example):

\begin{itemize}
   \item Sum of the number of math terms in an entire discussion dialog

   \item Mean of the number of math terms in an entire discussion dialog

   \item Variance in the number of math terms in an entire discussion dialog

\end{itemize}

\begin{figure*}[!t]
\vspace{-0.3cm}
\centering
  \includegraphics[width=7.0in]{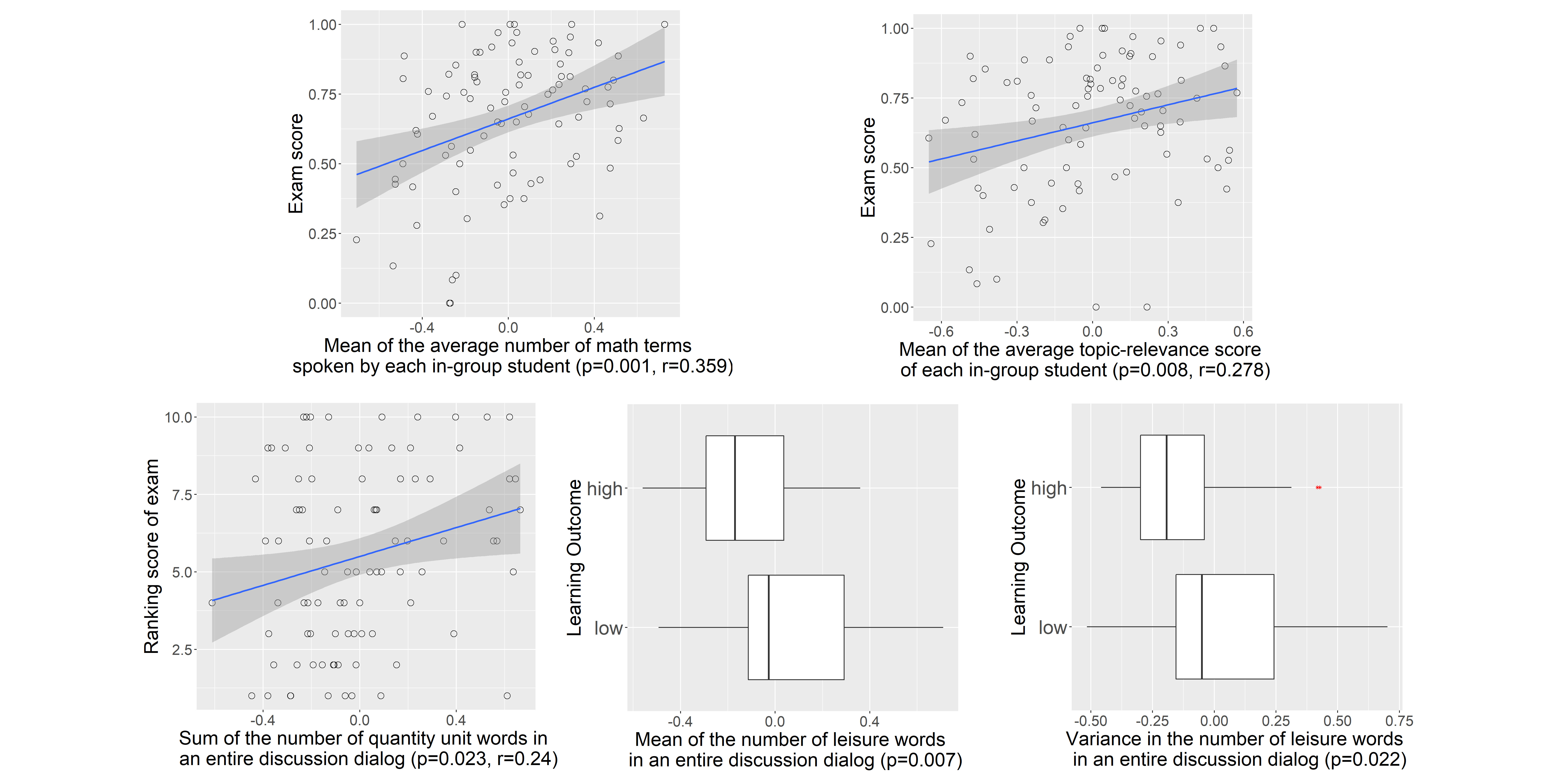}
\vspace{-0.3cm}
\caption{Results of statistical analysis between group learning outcome and features that represent topic relevance of discussions. A dot in scatter plot represents a data sample, which is one week's discussion of one group. The shading is the confidence interval of correlation analysis with confidence level of 95\%. A box in the box plot represents the degree of dispersion and skewness of the data in one category. Minimum, first quartile, median, third quartile and maximum of the data in one category are depicted from left to right in a box.}
\label{topic}     
\end{figure*}

\vspace{-0.1cm}
\section{Statistic Analyses between Dialog Features and Group Learning Outcome}
\label{stat}
\subsection{Group Learning Outcome Assessment}
\label{assess}
We chose to use the midterm and final examination scores as measurement of group learning outcomes.  The curriculum was carefully designed in such that each week contains a set of unique topics, and exact questions from the midterm and final examinations can be identified for each topic. For example, topics in Week 9 are covered by questions 10 and 15 in the final examination.  Such a mapping enables us to extract the grades of the members of a group on particular questions, in order to assess the group's learning outcome for the topic of a given week.  Considering that the grades of the midterm and final examinations respectively accounts for 20\% and 30\% of the final grade for each student, we decided to use proportional weighting to compute the learning outcome for a given student group on the topic of a specific week, as shown in Equation (\ref{E_s}). 
\begin{equation}
  E_s=\frac{0.4S_m+0.6S_f}{0.4F_m+0.6F_f} \label{E_s} 
\end{equation} 
where $S_m$ represents the grades earned by the group members from the midterm exam questions for the topic of a specific week, $S_f$ represents the grades earned by the group members from the final exam questions for the topic of a specific week, the full marks of $S_m$ is represented by $F_m$, and full marks of  $S_f$ is represented by $F_f$.

Since we are able to identify the group learning outcome for the topics covered in each week, we are also able to use the scores of the group to rank their relative learning outcome in each week. Since there are 10 groups in total, we can also give them a ranking score as 10(top) to 1(bottom). The ranking score comes in handy for comparison especially when the groups have very close scores for specific topics in a week.

\vspace{-0.1cm}

\subsection{Statistical Analyses}
\label{Exp}
In order to explore the indicators of group learning outcomes from group discussions, correlation analysis and ANalysis Of VAriance (ANOVA) are performed between group-level features and group learning outcomes.

\subsubsection{Correlation Analysis}
\label{cor}
\begin{itemize}
    \item Data
    
    As mentioned above, we are able to derive the group learning outcome (through exam scores and ranking scores) for topics that are covered for each week. We have a total of 9 weeks of discussion data (after the add/drop period until the end of the semester) for 10 groups of students. Thus we have transcriptions for 90 group discussions, which form the basis of the correlation analysis.

    \item Independent Variables
    
    The group-level discussion features are treated as independent variables. We note that there is variability across the weekly topics. For example, a more fundamental topic may not have too many new math terms, as compared with a more advanced topic, and consequently there would be markedly more math terms spoken for advanced topics. In order to normalize for similar effects, we applied min-max normalization among all groups in each week to scale all group-level features from -1 to 1. After normalization, each group-level feature in one week maintains the same relative magnitudes across groups, and each of the group-level feature across different weeks has the same feature scale. This step facilitates comparison across different groups and across different weeks.

    \item Dependent Variables
    
    As mentioned in the previous section, we are able to extract the scores of relevant questions in the midterm and final exams and map them to the corresponding topics covered, as well as the week they were covered. We also used ranking scores as a presentation. These scores that measure group learning outcome are adopted as dependent variables.

\end{itemize}

\subsubsection{ANalysis Of VAriance (ANOVA)}
\begin{itemize}
    \item Data
    
    ANOVA is performed to explore indicators that could significantly distinguish between groups with high versus low performance on learning outcomes. The top three groups of each week are labeled as \textit{High Learning Outcome}, whereas the bottom three groups of each week are labeled as \textit{Low Learning Outcome}. As mentioned in Sect.~\ref{cor}, we have a total of 9 weeks of discussion data. Therefore, each category contains 27 group discussions, which makes up the 54 analyzed discussion data in total.

    \item Independent Variables
    
    The independent variables in ANOVA are two categories -- \textit{High Learning Outcome} and \textit{Low Learning Outcome}, which is defined above.

    \item Dependent Variables
    
     All of the group-level features are adopted as dependent variables in ANOVA, which takes the same normalization as the correlation analysis. ANOVA is conducted on every group-level feature in order to examine whether it contributes significantly towards a \textit{High Learning Outcome} versus \textit{Low Learning Outcome}
     
\end{itemize}

\begin{figure*}[!t]
\centering
\vspace{-0.3cm}
  \includegraphics[width=7.0in]{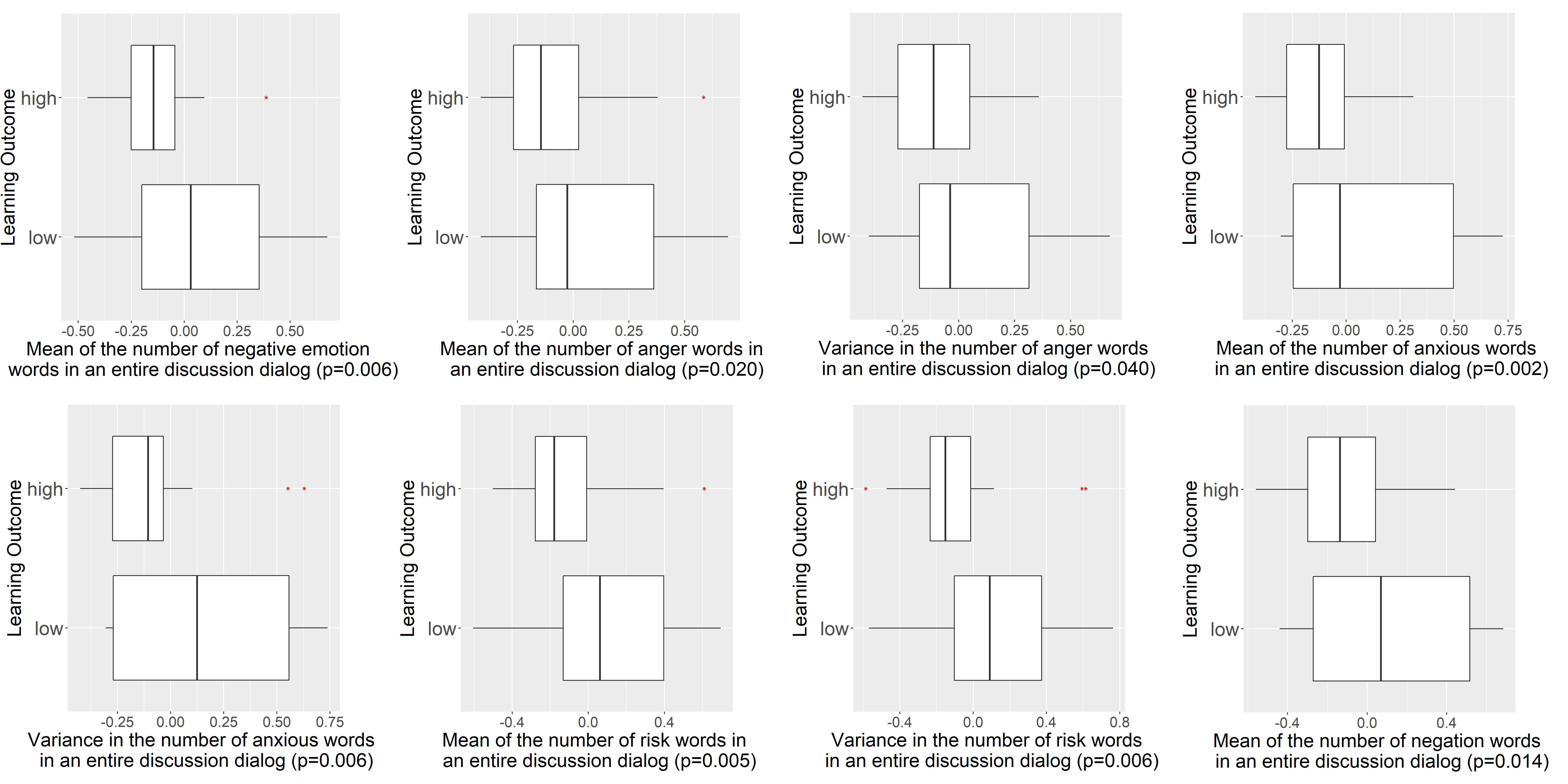}
\vspace{-0.3cm}
\caption{Results of statistical analysis between group learning outcome and features that indicate negative sentiment. A box in the box plot represents the degree of dispersion and skewness of the data in one category. Minimum, first quartile, median, third quartile and maximum of the data in one category are depicted from left to right in a box.}
\label{emotion}     
\vspace{-0.3cm}
\end{figure*}

\begin{figure}[!t]
\centering
  \includegraphics[width=3.4in]{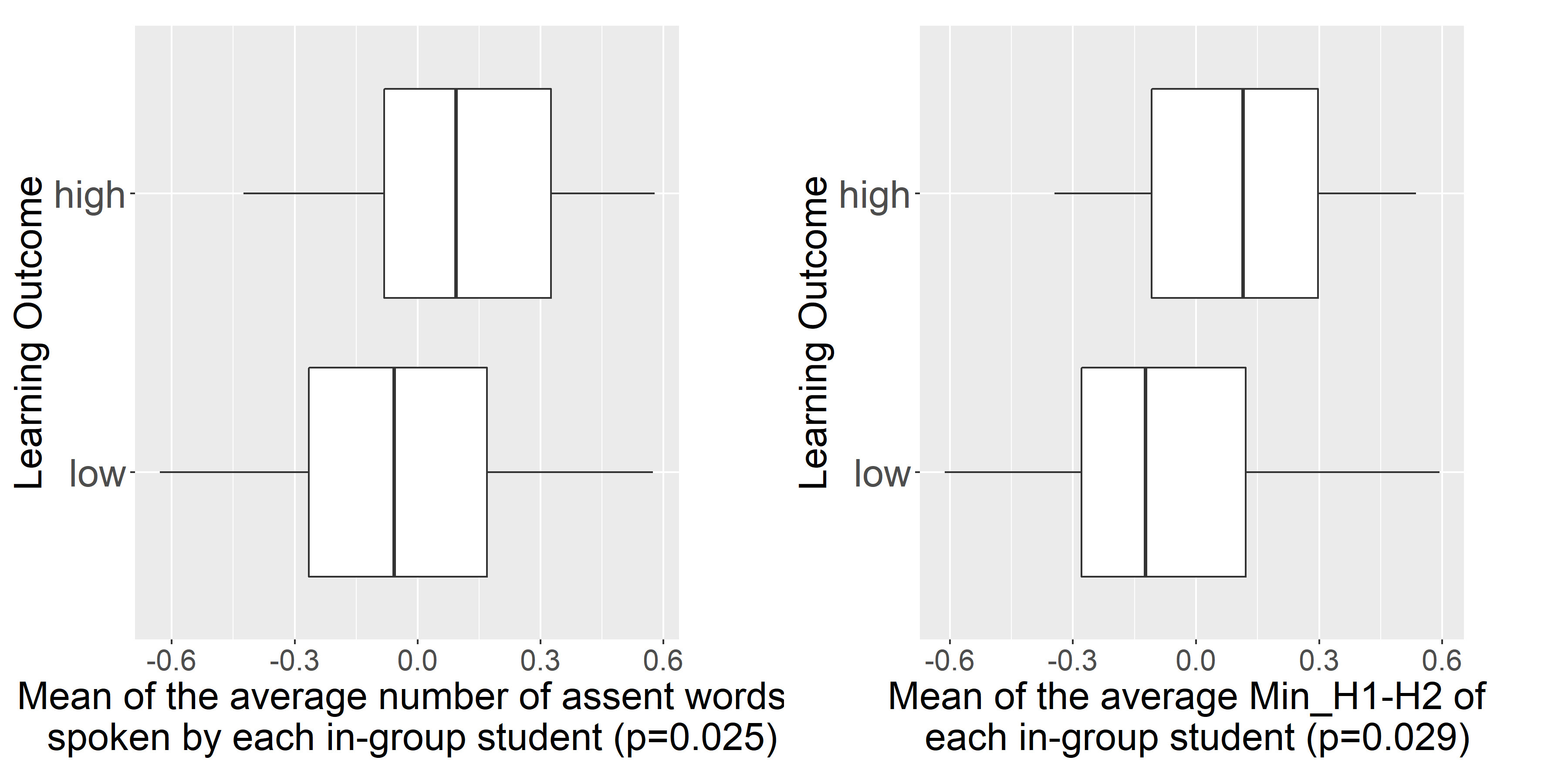}
\vspace{-0.3cm}
\caption{Results of statistical analysis between group learning outcome and features that indicate positive sentiment. A box in the box plot represents the degree of dispersion and skewness of the data in one category. Minimum, first quartile, median, third quartile and maximum of the data in one category are depicted from left to right in a box.}
\vspace{-0.3cm}
\label{emotion2}     
\end{figure}

\begin{figure*}[!t]
\centering
\vspace{-0.3cm}
  \includegraphics[width=7.0in]{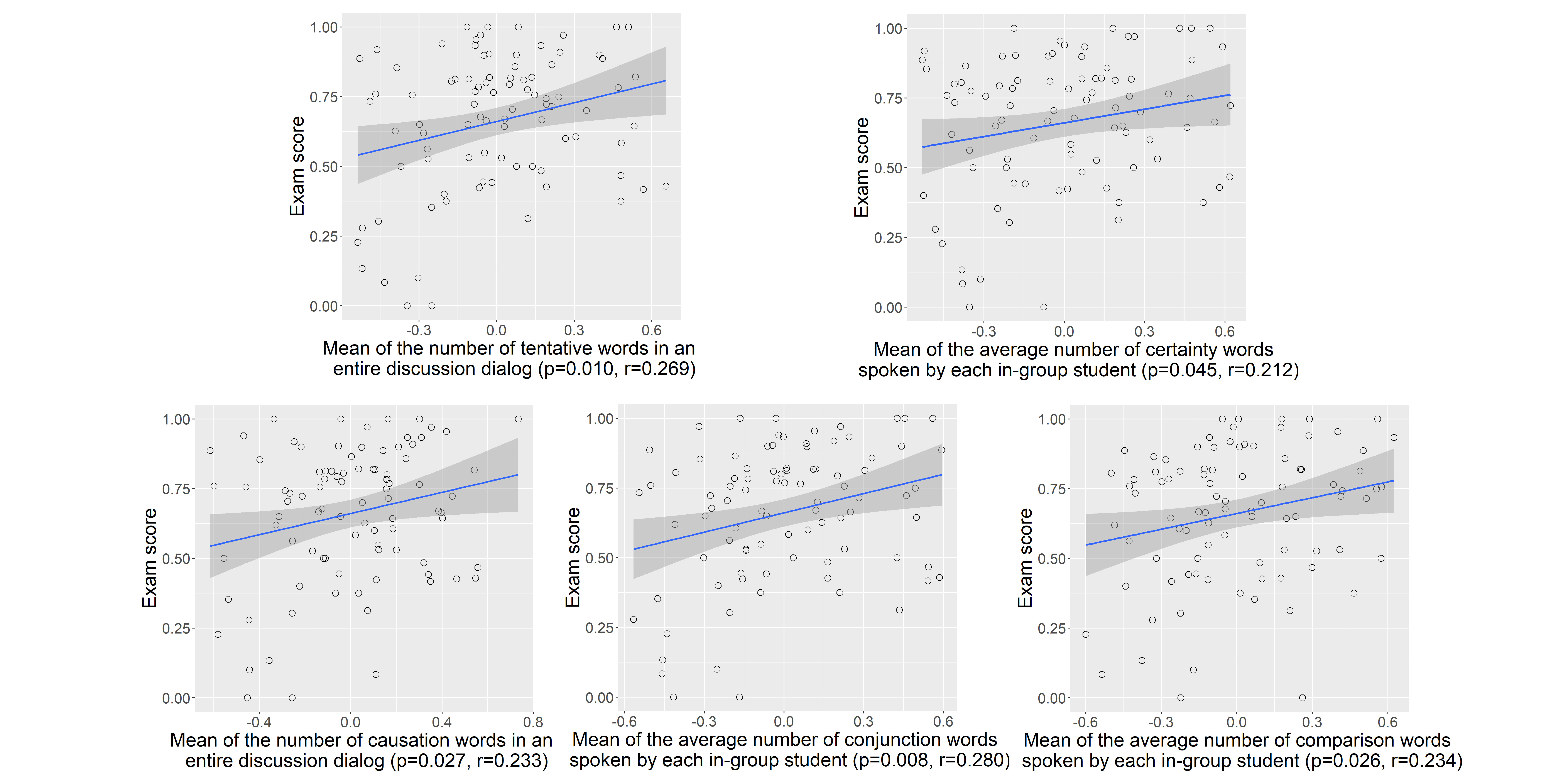}
\vspace{-0.3cm}
\caption{Results of statistical analysis between group learning outcome and features that reflect cognitive processing. A dot in scatter plot represents a data sample, which is one week's discussion of one group. The shading is the confidence interval of correlation analysis with confidence level of 95\%.}
\label{logical}
\vspace{-0.3cm}
\end{figure*}

\vspace{-0.1cm}
\subsection{Results and Discussion}
In this section, we will highlight the key results from statistical analyses. Results from correlation analysis will be presented by scatter plot, where the y-axis is the measurement of learning outcome and the x-axis is the normalized group-level features. Pearson correlation coefficient (r-value) and p-value are also provided below each sub-figure. ANOVA results will be presented by box plots, where the y-axis is group learning outcome and the x-axis is the normalized group-level features. The p-value is also provided below each sub-figure. In this work, we conducted statistical analysis at the significance level $\alpha = 0.05$.

\subsubsection{Topic Relevance}
First, we find that topic relevance of group discussions are important indicators that reflect group learning outcome. Fig. \ref{topic} shows some results of statistical analysis between group learning outcome and features that represent topic relevance of discussions. For the features with more than one statistical significant group-level features that reflect similar analysis results, we select to show the analysis result with the lowest p-value. As mentioned in Sect.~\ref{topic-sec}, the number of math terms and topic relevance score can represent the relevance between the discussion dialogs and topic. Besides, as mentioned in Sect.~\ref{social}, ``quantity unit'' category in LIWC dictionaries can detect the measure words that usually occur together with math-related words. Therefore, the number of quantity unit words could indicate topic relevance of group discussions. As shown in Fig. \ref{topic}, the mean of average number of math terms spoken by each in-group student, the mean of average topic relevance score of each in-group student, and the sum of the number of quantity unit words in an entire discussion dialog are positively correlated with the group learning outcome.  Moreover, the ``leisure" category in LIWC dictionaries could detect the leisure words that indicate students are relaxing for enjoyment rather than discussing the math-related topic. We observe from the ANOVA result that low learning outcome groups tend to have higher mean number of leisure words spoken in an entire discussion dialog, which suggests students in low learning outcome groups tend to have more relaxing off-topic discussion rather than on-topic discussion. We also find that high learning outcome groups tend to have lower variance in the number of leisure words in an entire discussion dialog, which suggests that high learning outcome groups tend to speak fewer leisure words. The topic relevance of in-class discussions should reflect the students' concentration in class. Higher concentration generally have higher learning outcomes, which is reflected in the results.

\subsubsection{Sentiment}
Sentiment is also an essential indicator that reflects group learning outcome. Fig. \ref{emotion} and Fig. \ref{emotion2} show some results of statistical analysis between group learning outcome and features that indicate sentiment. For the features with more than one statistical significant group-level features that reflect similar analysis results, we select to show the analysis result with the lowest p-value. The ``negative emotion", ``anger", ``anxiety", ``risk" and ``negation" categories in LIWC dictionaries (mentioned in Sect.~\ref{social}) could indicate the negative sentiment. The ``anger" and ``anxiety" category in LIWC dictionaries detect words that reflect angry and anxious emotion, which are both negative sentiment. The ``risk" category in LIWC dictionaries could detect the words that indicate something unpleasant or unwelcoming is likely to happen, which usually suggests a negative sentiment. The ``negation" category in LIWC dictionaries could detect the contradiction or denial words, which usually contain negative sentiment.  The ANOVA results show that the mean number of negative emotion/anger/anxiety/risk/negation words spoken in an entire discussion dialog are significantly greater in low learning outcome groups than in high learning outcome groups. Besides, we also find that the variance of number of anger/anxiety/risk words in an entire discussion dialog are all significantly lower in high learning outcome groups, which indicates that high learning outcome groups tend to insist on speaking a low number of negative sentiment words. In addition, the ``assent" category in LIWC dictionaries detects the words that indicate the expression of approval or agreement, which implies positive sentiment. High Min\_H1-H2 (mentioned in Sect.~\ref{acoustic}) indicates relief emotion, whereas low Min\_H1-H2 indicates fear or panic emotions \cite{patel2011mapping}. As shown in Fig. \ref{emotion2}, the mean of average number of assent words spoken by each in-group student, and the mean of average Min\_H1-H2 of each in-group student are significantly higher in high learning outcome groups than low learning outcome groups.  Overall, groups with high learning outcome tend to have more positive sentiment, whereas low learning outcome groups tend to have more negative sentiment.

\subsubsection{Cognitive Processing}
Fig. \ref{logical} shows some results of statistical analysis between group learning outcome and features that reflect cognitive processing. For the features with more than one statistical significant group-level features that reflect similar analysis results, we select to show the analysis result with the lowest p-value. As mentioned in Sect.~\ref{social}, ``tentative", ``certainty", ``causation", ``conjunction" and ``comparison" categories in LIWC dictionaries could measure the cognitive processing. The ``tentative" category could detect the words that indicate intention to try something, which to some extent could reflect students' curiosity in learning. We can observe from Fig. \ref{logical} that the mean of the number of tentative words in an entire discussion dialog shows significantly positive correlation with group learning outcome. The ``certainty" category in LIWC dictionaries aims to detect the words that show conviction about something. Fig. \ref{logical} shows that the mean of the average number of certainty words spoken by each in-group student is positively correlated with group learning outcome, which indicates that higher learning outcome groups tend to have higher conviction in discussion. The ``causation" category aims to detect the words that reflect the occurrence of relationship between cause and effect in discussions. The ``conjunction" category aims to detect the words that reflect the occurrence of coordinative logic relation in discussions. The ``comparison" category aims to detect the words that reflect the consideration of the similarities or dissimilarities between two things. These three categories in LIWC dictionaries can reflect the degree of logical thinking in discussions and more occurrences of these words indicate more logical thinking in discussions. Correlation analysis results show that the mean of the number of causation words in an entire discussion dialog, the mean of the average number of conjunction words spoken by each in-group student, the mean of the average number of comparison words spoken by each in-group student are all positively correlated with group learning outcome. The result supports that higher learning outcome groups tend to have more logical thinking in discussions.

\begin{figure}[!t]
\centering
  \includegraphics[width=3.4in]{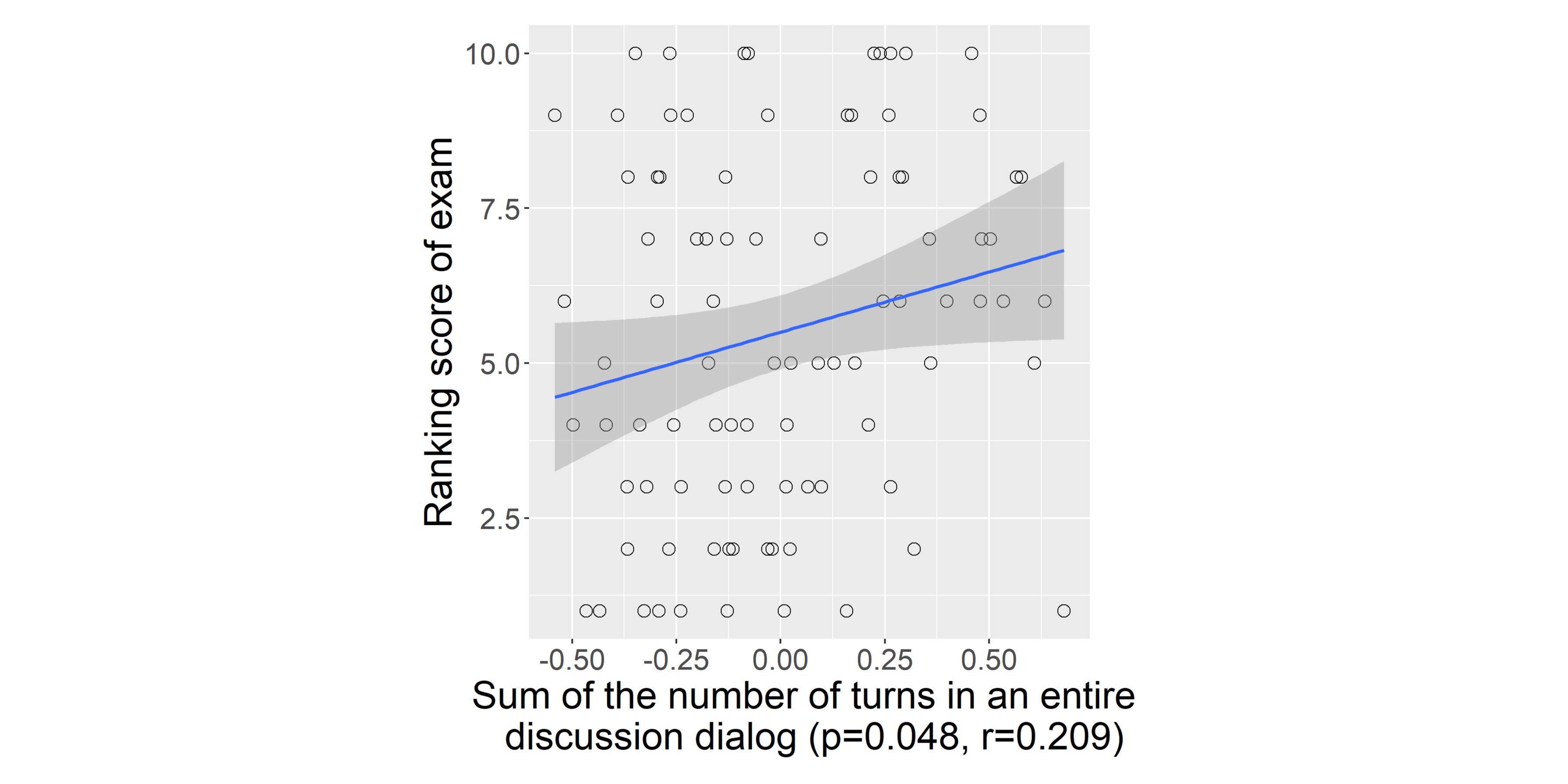}
\vspace{-0.3cm}
\caption{Results of statistical analysis between group learning outcome and the feature that reflects information exchange. A dot in scatter plot represents a data sample, which is one week's discussion of one group. The shading is the confidence interval of correlation analysis with confidence level of 95\%.}
\label{information}    
\vspace{-0.3cm}
\end{figure}

\begin{figure}[!t]
\centering
  \includegraphics[width=3.4in]{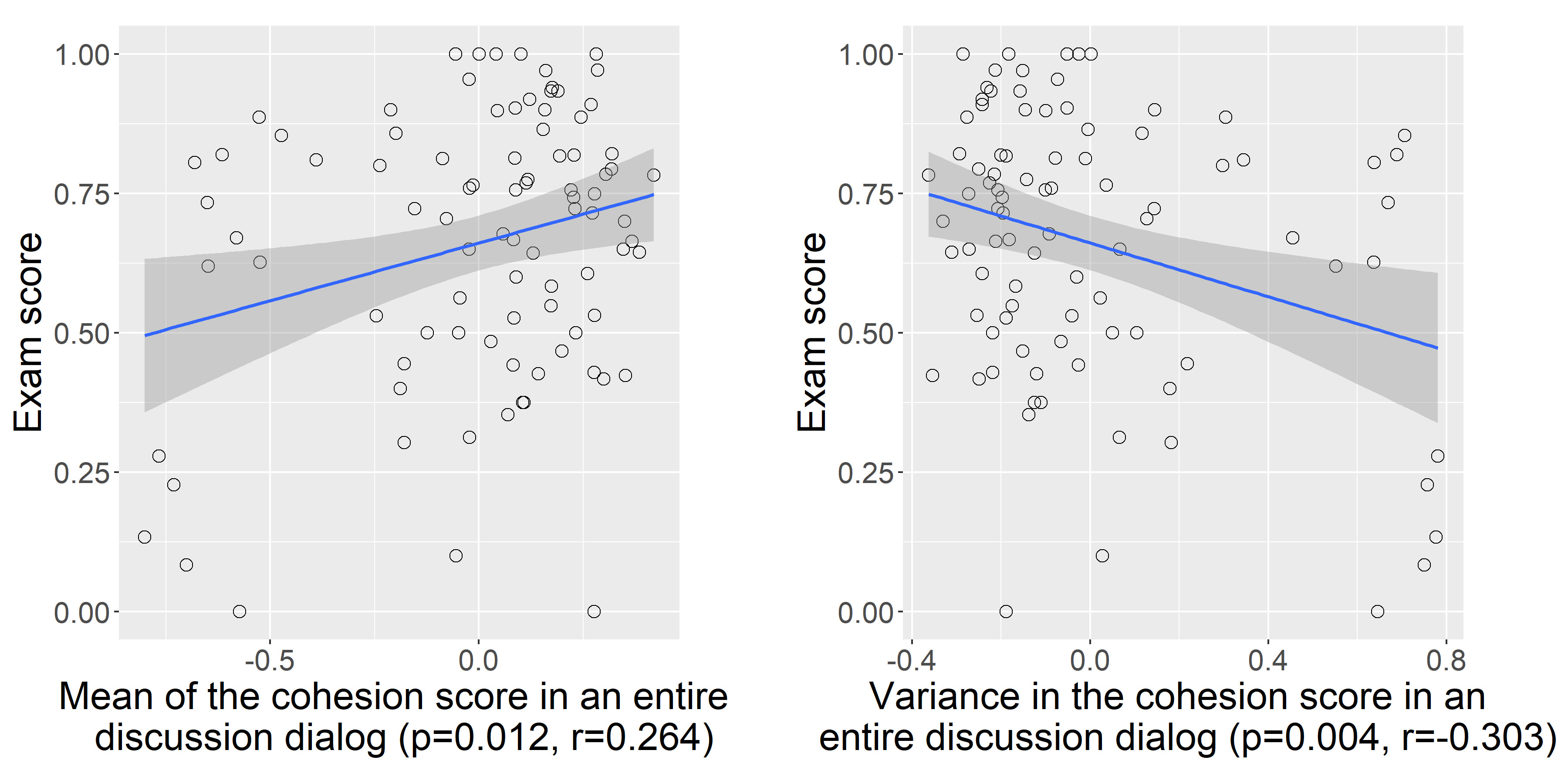}
\vspace{-0.3cm}
\caption{Results of statistical analysis between group learning outcome and the feature that measures context relevance in discussions. A dot in scatter plot represents a data sample, which is one week's discussion of one group. The shading is the confidence interval of correlation analysis with confidence level of 95\%.}
\label{cohesion}    
\vspace{-0.3cm}
\end{figure}

\subsubsection{Information Exchange}
The number of dialog turns can reflect the degree of information exchange during discussions. Fig. \ref{information} shows that the sum of the number of turns in a discussion dialog is positively correlated with the group learning outcome. This indicates that the degree of information exchange during discussions positively correlates with the group learning outcome.

\subsubsection{Context Relevance}
Fig. \ref{cohesion} shows some results of statistical analysis between group learning outcome and the feature that measures context relevance in discussions. For the statistical significant results with similar trends, we select to show the analysis result with the lowest p-value. As mentioned in Sect.~\ref{context}, the cohesion score is created to measure the context relevance of the discussion. Higher content similarity between two neighbouring segments leads to a higher cohesion score. Correlation analyses show that the mean of the cohesion score in an entire discussion dialog is positively correlated with the group learning outcome, which indicates that high context relevance discussions are associated with higher group learning outcome. Moreover, the variance of the cohesion score in an entire discussion dialog is negatively correlated with the group learning outcome, which indicates that stably high context relevance of discussions is positively correlated with group learning outcome.

\vspace{-0.1cm}
\section{Automatic Prediction of Group Learning Outcome from Discussion Dialog Features}
\label{ML}
Upon identifying the proper features from group discussions that reflect the group learning outcome, automatic prediction becomes possible. In this work, different machine learning algorithms are used to predict the group learning outcome based on statistically significant group-level features. First, we label the top three, middle four and bottom three groups based on learning outcomes of every week's topics as \textit{High Learning Outcome}, \textit{Mid Learning Outcome} and \textit{Low Learning Outcome} respectively. We have nine weeks of discussion data, thus our data consist of 27 \textit{High Learning Outcome} discussions, 36 \textit{Mid Learning Outcome} discussions and 27 \textit{Low Learning Outcome} discussions (90 samples in all). It is understood that the experimental setup has limited data and the collection of group discussion data is challenging in both social and technical perspective. Although the experimental data is limited, we already have much more data than previous discussion analysis researches, where no more than 60 labelled group discussion data samples are collected \cite{kubasova2019analyzing, avci2016predicting, murray2018predicting, ochoa2013expertise, scherer2012multimodal, martinez2013capturing, reilly2019predicting, spikol2017estimation}. Then we select statistically significant group-level features based on statistical analyses. Sixty two features are selected. In order to further prevent overfitting, random projection \cite{bingham2001random} is applied to do the dimension reduction, which reduces the feature dimension from 62 to 45. Naive Bayes (NB), Neural Network (NN), K-Nearest Neighbours (KNN), Random Forest (RF), LightGBM (LGBM) \cite{NIPS2017_6907}, XGBoost (XGB) \cite{chen2016xgboost} and Support Vector Machine (SVM) are used to automatically classify the group learning outcome into \textit{High Learning Outcome}, \textit{Mid Learning Outcome} and \textit{Low Learning Outcome}.

We perform five-fold cross-validation in prediction, which takes the prediction five times with five different subsets of test data and then take the average of prediction results. In each fold, data is divided into a training set and a test set with the ratio of 4:1. Then the training set is subdivided into a sub-training set and a validation set for five times, and collections of model hyper-parameters (e.g. value of regularization parameter $C$ and kernel coefficient $\gamma$ in SVM) which are selected by Bayes optimization algorithm \cite{snoek2012practical} are compared according to the average of five validation scores. We select the hyper-parameters with the best average validation score as the model hyper-parameters. Then the model is retrained on the whole training set using the selected hyper-parameters. Finally, the trained model is applied on the test set to get one out of five fold test score.



\begin{figure}[!t]
\centering
\vspace{-0.2cm}
\includegraphics[width=2.8in]{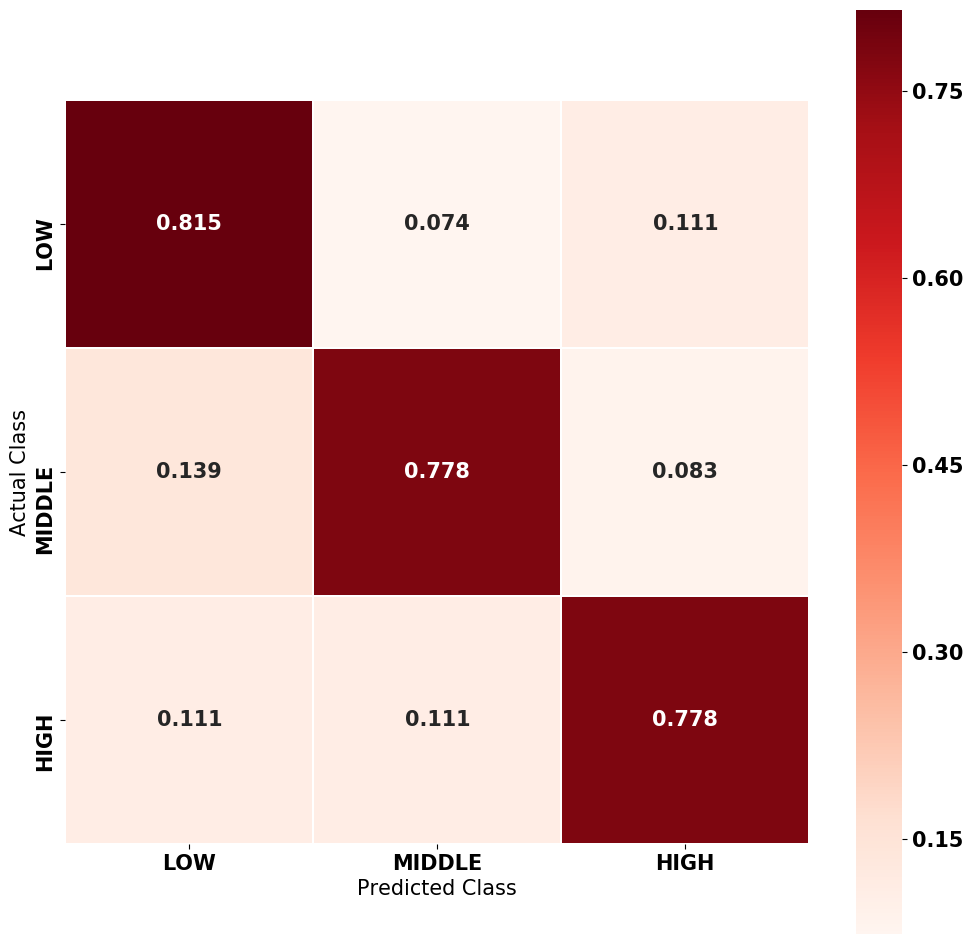}
\vspace{-0.2cm}
\caption{Confusion matrix from the prediction result of SVM model}
\label{svm_e}   
\vspace{-0.4cm}
\end{figure}

In each fold, the prediction accuracy is calculated according to the equation (\ref{eq_acc}), 
\begin{equation}
\vspace{-0.1cm}
 ACC=\frac{N_{match}}
  {N_{all}}
  \label{eq_acc}
\end{equation}
where $N_{match}$ represents the number of prediction results that match ground-truth labels, and $N_{all}$ represents the total number of test data in each fold. SVM model reaches the best prediction accuracy of 78.9\%, which demonstrates the feasibility of achieving automatic group learning outcome prediction based on group discussion dialog in flipped classroom. The prediction results vastly outperform the random guess (33.3\%), which indicates that our model derive meaningful classification from selected features.  We also depict the confusion matrix from the prediction result of SVM model. As shown in Fig. \ref{svm_e}, three classes are classified correctly in balance, only \textit{Low learning outcome} has reached marginally higher recall than the other two classes.


\section{Conclusion}
This paper present a novel study on the correlation between student in-class discussion dialog and group learning outcomes in the flipped classroom. We collected a new Chinese and English code-switched flipped classroom audio corpus throughout a semester with stable student enrollment, and the group discussion audio was separated from the lecture audio by a customized speech classification technology. The group discussion audio recordings have been hand-transcribed with speaker diarization annotations in order to facilitate analysis. Some proper tools and customized technical processing frameworks are introduced to extract the spoken dialog features from bilingual dataset. Several important indicators from discussion dialog that reflect the group learning outcome were identified through statistical analysis. We found that topic relevance of discussions, positive sentiment in discussions, curiosity in learning, certainty in discussions, degree of logical thinking in discussions, information exchange during discussions and context cohesion of discussions are positively correlated with group learning outcome. Finally, machine learning algorithms were given statistically significant indicators to automatically classify the group learning outcome into high, middle or low. Best classification result reached the accuracy of 78.9\%, which not only showed that the explored indicators did contribute to reflect the group learning outcome, but also demonstrated the feasibility of achieving automatic learning outcome prediction from group discussion dialog in flipped classroom.


\ifCLASSOPTIONcaptionsoff
  \newpage
\fi



\bibliographystyle{IEEEtran}
\bibliography{ref.bib}
\end{document}